\documentclass[10pt,conference]{IEEEtran}
\IEEEoverridecommandlockouts
\usepackage[style=ieee,url=false,doi=false,isbn=false]{biblatex}
\usepackage{amsmath,amssymb,amsfonts}
\usepackage{algorithm}
\usepackage{algpseudocode}
\usepackage{graphicx}
\usepackage{textcomp}
\usepackage{xcolor}
\usepackage{braket}
\usepackage{physics}
\usepackage{subfigure}
\usepackage{bm}

\def\BibTeX{{\rm B\kern-.05em{\sc i\kern-.025em b}\kern-.08em
    T\kern-.1667em\lower.7ex\hbox{E}\kern-.125emX}}

\begin{document}

\title{
Sampling-based Quantum Optimization Algorithm with Quantum Relaxation
  
  \thanks{
    This work was performed for Council for Science, Technology and Innovation (CSTI), Cross-ministerial Strategic Innovation Promotion Program (SIP), ``Promoting the application of advanced quantum technology platforms to social issues'' (Funding agency: QST).
  }
}

\makeatletter
\newcommand{\linebreakand}{%
\end{@IEEEauthorhalign}
\hfill\mbox{}\par
\mbox{}\hfill\begin{@IEEEauthorhalign}
}
\makeatother

\author{
  \IEEEauthorblockN{
    Hiromichi Matsuyama\IEEEauthorrefmark{1},
    and
    Yu Yamashiro\IEEEauthorrefmark{1}
  }
  \IEEEauthorblockA{
    \IEEEauthorrefmark{1}\textit{Jij Inc.}, 3-3-6 Shibaura, Minato-ku, Tokyo, 108-0023, Japan
  }

  \IEEEauthorblockA{
    h.matsuyama@j-ij.com,
    y.yamashiro@j-ij.com,
  }
}

\maketitle

\begin{abstract}
Combinatorial optimization is one of the promising applications of quantum computation.
Variational Quantum Algorithm (VQA) is a classical-quantum hybrid algorithm framework for noisy quantum devices.
However, the quality of the solution is readily affected by statistical fluctuations and physical noise, making it challenging to maintain applicability for large-scale problems.
In contrast, sampling-based quantum algorithms have recently been successfully applied to large-scale quantum chemistry problems.
In the sampling-based framework, the quantum device is used only for sampling, while the ground state and its energy are estimated on the classical device.
In this study, we propose the Sampling-based Quantum Optimization Algorithm (SQOA) that leverages these advantages for combinatorial optimization.
There are two challenges in constructing SQOA.
The first challenge is that we need to encode the optimization problem in a non-diagonal Hamiltonian, even though many VQAs encode it into the Ising Hamiltonian, which is diagonal. 
The second challenge is that we need a method for efficient preparation of the input state to be sampled.
To address the first challenge, we utilize the Quantum Relaxation (QR) method for encoding an optimization problem within a Hamiltonian.
This method achieves a higher encoding density by mapping multiple variables to a single qubit, thus requiring fewer qubits for larger problems compared to the Ising Hamiltonian approach.
Moreover, to address the second challenge, we investigate the parameter transferability within the Quantum Alternating Operator Ansatz for QR Hamiltonians.
We show that restricting parameters to a linear form exhibits moderate transferability for 3-regular MaxCut problems, similar to transferability observed in the Quantum Approximate Optimization Algorithm.
Therefore, this property allows us to efficiently prepare the input state for a large problem instance using the parameters obtained from a small problem instance.
We leveraged transferability to create input states and applied SQOA with QR to the MaxCut instances.
By transferring parameters from a 20-node problem, we demonstrate that SQOA with QR provides high-quality solutions for 40-node problems without variational parameter optimization.
\end{abstract}


\section{Introduction}
\label{sec:intro}
Combinatorial optimization involves finding solutions that minimize the objective value within a large solution space~\cite{Laurence_Wolsey2020-gj, korte2011combinatorial, conforti2014integer}.
Various real-world problems can be formulated as combinatorial optimization problems, including vehicle routing problems~\cite{ralphs2003capacitated} and unit commitment problems~\cite{padhy2004unit}.
Combinatorial optimization is one of the promising applications for quantum computation~\cite{Abbas2024challenge}.
Variational Quantum Algorithm (VQA) is a quantum-classical hybrid algorithm framework for Noisy Intermediate Scale Quantum (NISQ) devices~\cite{Cerezo2021-zl, bharti2022nisq}.
Quantum Approximate Optimization Algorithm (QAOA) and its extensions are VQAs designed for optimization problems~\cite{Farhi2014-lj, BLEKOS2024review}.
While VQAs hold promise for NISQ devices, the estimation of expectation values on actual quantum hardware renders them susceptible to errors arising from statistical fluctuations and physical noise.
Besides the issue of physical noise, the optimization of variational parameters faces the barren plateau problem~\cite{McClean2018barren, larocca2024reviewbarrenplateausvariational}.

Quantum-Selected Configuration Interaction (QSCI), a non-variational sampling-based quantum algorithm for calculating ground states using NISQ devices, has been proposed in quantum chemistry~\cite{kanno2023qsci}.
Unlike VQAs, QSCI only performs sampling on quantum devices.
The sampling results are used to construct an effective Hamiltonian, which is then diagonalized on a classical device to obtain the ground state and its energy.
Since ground state energy is estimated on classical devices, QSCI is less susceptible to physical noise than VQAs.
QSCI has already been successfully applied to large-scale quantum chemistry problems on real hardware, providing good approximate solutions~\cite{Robledo-Moreno2024-sl}.

Sampling-based algorithms are a promising new technique for noisy devices.
However, two challenges arise when applying sampling-based algorithms to optimization problems.
The first issue is that QSCI is designed for Hamiltonians with non-diagonal components.
Optimization problems are usually encoded in an Ising Hamiltonian, which is diagonal.
Thus, applying sampling-based approaches requires designing a non-Ising Hamiltonian to encode the optimization problem.
The second issue is that QSCI requires an input state that roughly approximates the target ground state for ground state energy estimation.
If we cannot prepare this state efficiently, the classical computational cost increases, leading to difficulties in applying this method to larger problems.

In this paper, we propose a Sampling-based Quantum Optimization Algorithm (SQOA) with Quantum Relaxation.
We refer to our proposed algorithm as SQOA-QR for brevity.
SQOA-QR is a non-variational quantum relaxation-based optimization algorithm that addresses both challenges mentioned above.
The non-variational nature of SQOA-QR provides inherent advantages for NISQ devices by reducing sensitivity to noise and eliminating the computational overhead of iterative parameter optimization. 
To address the first challenge concerning problem Hamiltonians, we encode optimization problems using quantum relaxed Hamiltonians derived from the Quantum Random Access Optimization (QRAO) algorithm~\cite{Fuller2021-su, Teramoto2023-ix, Teramoto2023-jd}.
This method utilizes the Pauli $X$ and $Y$ operators in addition to the Pauli $Z$ operator used in the Ising Hamiltonian to encode classical variables.
As a result, it is possible to achieve a compression ratio up to three times higher than the Ising Hamiltonian.
This higher compression ratio reduces the number of required qubits.

To address the second challenge related to the preparation of the input state for QSCI, we draw upon insights from recent work~\cite{He2025-br}.
He \textit{et al.} applied the framework of the Quantum Alternating Operator Ansatz~\cite{Hadfield2019-lb, kremenetski2021quantumalternatingoperatoransatz} to QRAO~\cite{He2025-br}, termed QAOA-for-QRAO.
They showed that the parameter setting heuristic proposed for QAOA is also applicable to QAOA-for-QRAO.
In their method, optimal variational parameters are precomputed for many problem instances with several problem sizes. 
The parameters are then estimated based on the precomputed optimal parameters and subsequently reused for diverse problem instances, including those of larger size.
The results indicate parameter transferability in QAOA-for-QRAO, suggesting that parameters can be transferred between different problem instances.

In this study, we apply a recently proposed method for QAOA to QAOA-for-QRAO to prepare the input state of QSCI. 
This method combines linearization of the variational parameters and parameter transfer~\cite{Sakai2024-li}.
We refer to this method as LINXFER.
LINXFER only requires four parameters, regardless of the number of repetition layers, as a result of linearization.
Since these parameters need optimization for only one instance due to the parameter transferability, LINXFER is less computationally demanding than previously used methods.
We have demonstrated that LINXFER for QAOA-for-QRAO outperforms direct optimization to QAOA-for-QRAO with parameter random initialization in 3-regular unweighted MaxCut problems.
Interestingly, parameters estimated for a 20-node problem are effective for 40-node problems as well.
In SQOA-QR, we performed QSCI using the state prepared by LINXFER as the input state.
As a result, we obtained a high-quality solution without variational parameter optimization.
Therefore, combining LINXFER and QSCI provides an efficient approach to solving MaxCut problems without optimizing variational parameters.

This paper is organized as follows.
First, in Sec.~\ref{sec:background}, we explain the background of our proposed algorithms, including QAOA, QRAO, and QSCI.
In Sec.~\ref{sec:linxfer} and Sec.~\ref{sec:qsci}, we discuss our proposed algorithms: LINXFER for QAOA-for-QRAO and SQOA-QR, respectively. 
For each algorithm, we first explain the details of the method and then present the experimental results to validate its effectiveness.
In Sec.~\ref{sec:summary}, we summarize the results and discuss the future direction of this study.
The Appendix demonstrates experimental results when using LINXFER parameters as warm start initial values for variational parameter optimization.

\section{Background \label{sec:background}}

\subsection{Quantum Approximate Optimization Algorithm}
The Quantum Approximate Optimization Algorithm (QAOA) is a variational quantum algorithm for solving combinatorial optimization problems~\cite{Farhi2014-lj}.
In QAOA, two parametrized unitary operators, the cost unitary~$e^{-i\gamma H_C}$ and the mixer unitary~$e^{-i\beta \sum_i X_i}$, are alternately applied to create a variational quantum circuit.
Here, $\beta$ and $\gamma$ are the variational parameters to be optimized, and $X_i$ is the Pauli $X$ operator acting on the $i$th qubit.
We encode the cost function of the optimization problem $C(\bm{z})$ into quantum Hamiltonian~$H_C$.
The variational quantum circuit for QAOA with $p$ iterative layers, containing $2p$ parameters, is
\begin{equation}
\ket{\psi(\bm{\beta}, \bm{\gamma})} = e^{-i\beta_p \sum_i X_i}e^{-i\gamma_p H_C}\cdots e^{-i\beta_1 \sum_i X_i}e^{-i\gamma_1 H_C}  \ket{\psi_0}
\end{equation}
where $\ket{\psi_0} = \ket{+}^{\otimes N} = \qty(\frac{1}{\sqrt{2}}\qty(\ket{0} + \ket{1}))^{\otimes N}$ is the initial equal superposition state.
In QAOA, the classical computer is used to determine the best parameters $\bm{\beta}^*, \bm{\gamma}^*$ to minimize the expectation value of the cost Hamiltonian, $\bra{\psi(\bm{\beta}, \bm{\gamma})}H_c\ket{\psi(\bm{\beta}, \bm{\gamma})}$.
Since state~$\ket{\psi(\bm{\beta}^*, \bm{\gamma}^*)}$ is expected to be in the ground state or near the ground state of $H_C$, at the final step of QAOA, the classical solutions are obtained by sampling this state.

We typically choose the Ising Hamiltonian as our cost Hamiltonian $H_C$ since the Quadratic Unconstrained Binary Optimization (QUBO) problem can be easily encoded into the Ising Hamiltonian.
For example, the MaxCut problem on an undirected graph $G(V, E)$, which is the optimization problem we deal with in this paper, can be written as
\begin{equation}
        \max_{\bm{s} \in \{1,-1\}^n} C(\bm{s}) =\max_{\bm{s} \in \{1,-1\}^n} \sum_{(i,j) \in E} \frac{1}{2}\qty(1 - s_is_j)
        \label{eq:MaxCut}
\end{equation}
where $n = \abs{V}$ is the number of nodes.
In the MaxCut problem, we divide $V$ into two subsets to maximize the number of edges with both ends in different subsets.
Equation~\eqref{eq:MaxCut} can be mapped into quantum Hamiltonian:
\begin{equation}
H_C = -\sum_{(i,j) \in E}\frac{1}{2} \qty(1 - Z_iZ_j)
\label{eq:MaxCut_hamiltonian}
\end{equation}
where $Z_i$ is the Pauli $Z$ operator acting on $i$th qubit.
In the Hamiltonian above, we multiply the entire expression by $-1$ to convert the maximization problem into a minimization problem.
We can solve the MaxCut problem by finding the ground state of Eq.~$\eqref{eq:MaxCut_hamiltonian}$.

QAOA requires classical variational parameter optimization, and this computational cost increases as $p$ increases.
Various variational parameter optimization methods have been proposed to reduce computational cost~\cite{zhou2020qaoa, Brandao2018-lg, Streif2020traning, wurtz2021fixedangle, galda2021transferability, Shaydulin2023parameter, montanezbarrera2024transferlearning, Shaydulin2021symmetry,kremenetski2021quantumalternatingoperatoransatz, Sakai2024-li, Zhang2025grover, montanezbarrera2024universal}.
Among these methods, there are two notable strategies.
The first strategy is parameter transfer~\cite{Brandao2018-lg, Streif2020traning, wurtz2021fixedangle, galda2021transferability, Shaydulin2023parameter, montanezbarrera2024transferlearning}.
In QAOA, it has been reported that optimal variational parameters only weakly depend on the problem instance and tend to concentrate around similar values~\cite{Brandao2018-lg, Streif2020traning, wurtz2021fixedangle, Sack2021quantumannealing}.
This behavior is referred to as parameter concentration~\cite{akshay2021parameter}.
Parameter transfer leverages this concentration phenomenon by precomputing concentrated parameters and reusing these parameters across different instances.
This parameter-reusing approach significantly reduces the computational cost.
The second strategy involves restricting parameters to specific forms~\cite{Shaydulin2021symmetry,kremenetski2021quantumalternatingoperatoransatz, Sakai2024-li, Zhang2025grover, montanezbarrera2024universal}.
This method constrains parameters, $\bm{\beta}$ and $\bm{\gamma}$, to follow specific functional forms dependent on $p$ and optimizes only the parameters within the functions.
Since the number of parameters in these restricted functions does not depend on $p$, the computational cost remains independent of $p$.
Common choices for these functional forms include linear and constant functions.
Furthermore, parameter concentration phenomena have been observed even with restricted parameter forms~\cite{Zhang2025grover}. 
Parameter transfer combined with parameter form restriction has also been shown to provide high-quality solutions~\cite{montanezbarrera2024universal, Sakai2024-li}.
In this study, we applied these methods to a quantum Hamiltonian that is not an Ising Hamiltonian and demonstrated its applicability.

Quantum Alternating Operator Ansatz (QAOAnsatz) is a generalization framework of QAOA~\cite{Hadfield2019-lb}.
QAOA is inspired by Quantum Annealing~\cite{kadowaki1998, Farhi2001-zv}, which constrains the mixer Hamiltonian to be the transverse field of the Ising Hamiltonian. 
In contrast, QAOAnsatz lifts this constraint, allowing for more flexibility in algorithm design.
In the framework of QAOAnsatz, we can design any form of the cost unitary~$U_C(\gamma)$ and mixer unitary~$U_M(\beta)$.
These parameterized unitary operators are alternately applied as follows:
\begin{equation}
    \ket{\psi(\bm{\beta}, \bm{\gamma})} = U_M(\beta_p)U_C(\gamma_p)\cdots U_M(\beta_1)U_C(\gamma_1) \ket{\psi_0}.
\end{equation}
In QAOAnsatz, the initial state~$\ket{\psi_0}$ is chosen based on the form of the mixer unitary operators.
In the previous studies, the mixer Hamiltonian has been designed to search only the feasible regime of the optimization problem~\cite{Hadfield2019-lb, wang2020xy}.

\subsection{Quantum Random Access Optimization}
QRAO~\cite{Fuller2021-su} is a quantum relaxation-based optimization method that utilizes Quantum Random Access Code (QRAC)~\cite{Ambainis1999-cg, Ambainis2002-sb} to construct a quantum relaxed Hamiltonian.
The $(3,1,p)$-QRAC is an encoding method for encoding three classical bits $x_1,x_2,x_3 \in \{0,1\}$ into a single qubit as
\begin{equation}
    \rho(\bm{x}) = \frac{1}{2}\qty(I + \frac{1}{\sqrt{3}}\qty((-1)^{x_1}X + (-1)^{x_2}Y + (-1)^{x_3}Z)).
\end{equation}
Here, three classical bits are encoded into the coefficients of the Pauli $X, Y$ and $Z$ operators.
We can decode $x_i$ with probability $p=\frac{1}{2}\qty(1 + \frac{1}{\sqrt{3}})\approx 0.79$.

QRAO has two steps: to create a quantum relaxed Hamiltonian~$\tilde{H}_C$ and to decode the classical solution from the low energy state of $\tilde{H}_C$.
We will explain how to implement $(3,1,p)$-QRAO from Eq.~\eqref{eq:MaxCut}, and our explanation is based on \cite{Fuller2021-su}.
In order to construct $\tilde{H}_C$, we first need to decide the mapping of the classical bits and qubits.
When we create this mapping, we must avoid encoding adjacent nodes in $G(V,E)$ in the same qubit.
If we encode the two variables $s_i, s_j$ in the operators $X_i^\ell$ and $Y^\ell_j$ acting on the same $\ell$th qubit, the interaction term $s_is_j$ becomes $X^\ell_iY^\ell_j \propto Z^\ell_k$.
Since $Z^\ell_k$ corresponds to another variable $s_k$, the structure of the problem becomes different from the original problem.
Therefore, the adjacent nodes must be encoded in different qubits to keep the problem structure.

In order to create a correct mapping, first, we need to solve the vertex coloring problem on $G(V,E)$.
For this purpose, the greedy algorithm is sufficient.
Next, the node $v_i$ contained in the node-set $V_c$ of color $c$ is assigned to a qubit.
Since we use three Pauli operators, we can assign at most three variables to a single qubit.
As a result, the quantum relaxed Hamiltonian corresponding to Eq.~\eqref{eq:MaxCut} becomes
\begin{equation}
\tilde{H}_C = - \sum_{(i,j) \in E} \frac{1}{2}\qty(1 - 3P^{q_i}_iP^{q_j}_j).
\label{eq:relax_hamiltonian}
\end{equation}
Here, $q_i$ is the qubit index corresponding to $s_i$ and $P^\ell_i \in \{X^\ell_i, Y^\ell_i, Z^\ell_i\}$ represents the Pauli operator acting on the $\ell$th qubit.
This encoding allows us to reduce the number of bits by up to three times compared to Eq.~\eqref{eq:MaxCut}.
$\tilde{H}_C$ has non-diagonal components, so its ground state, $\tilde{\rho}$, is a quantum state and does not match the optimal solution of the original problem.
However, the ground state energy provides a lower bound of the original minimization problem.
In this section, we have explained only $(3,1,p)$-QRAO; however, $(2,1,p)$-QRAO~\cite{Fuller2021-su} and $(3,2,p)$-QRAO~\cite{Teramoto2023-ix} have also been proposed.

The second step is to obtain a classical solution from the obtained quantum state, $\rho$.
In this study, we used Pauli rounding in our experiments, so we explain this decoding heuristic method here.
In Pauli Rounding, we decode the classical variable $s_i$ from $\Tr[P^{q_i}_i \rho]$  as
\begin{equation}
    s_i = \mathrm{sgn}\qty(\Tr[P^{q_i}_i \rho]).
\end{equation}
Note that several rounding algorithms have been proposed~\cite{Fuller2021-su, Kondo2025-dd}.
However, in most cases, Pauli Rounding provides better solutions than magic state rounding in experiments~\cite{Teramoto2023-jd, Fuller2021-su}.

\subsection{Quantum Selected Configuration Interaction}
QSCI is a sampling-based quantum-classical hybrid algorithm for calculating eigenvalues and eigenstates of quantum Hamiltonians~\cite{kanno2023qsci, nakagawa2024adapt-qsci, Robledo-Moreno2024-sl}.
QSCI is designed to calculate not only for the ground state but also for the excited states~\cite{kanno2023qsci}. 
However, we only discuss the ground state here.
In this method, the ground state and its energy are obtained by diagonalizing the effective Hamiltonian on a classical device.
This effective Hamiltonian is constructed using the bitstrings obtained by measuring the input state on a computational basis.
This method has provided good approximate solutions for large-scale quantum chemistry problems~\cite{Robledo-Moreno2024-sl}.

We consider the problem of finding the approximate ground state of a $N$-qubit quantum Hamiltonian $H$.
In QSCI, we first prepare the \textit{input state}~$\ket{\Psi}$ on the quantum device.
Next, we repeatedly measure the input state in the computational basis.
The outcome is the set of the bitstrings $\bm{x} \in \{0,1\}^N$.
We take $R$ most frequent computational basis states from the set and construct the subspace $\mathcal{S}_R$ as 
\begin{equation}
\mathcal{S}_R = \mathrm{span}\{\ket{\bm{x}_1},\dots,\ket{\bm{x}_R}\}.
\end{equation}
We define the projection operator $P_R = \sum_{\mathcal{S}_R}\ket{\bm{x}}\bra{\bm{x}}$ onto subspace $\mathcal{S}_R$.
The effective Hamiltonian restricted to subspace $\mathcal{S}_R$ is
\begin{equation}
H_R = P_R H P_R.
\label{eq:effective_hamiltonian}
\end{equation}
Then, the ground state~$\ket{\psi_R}$ and its energy~$E_R$ of this $H_R$ are calculated on the classical devices, and QSCI method outputs $\ket{\psi_R}$ and $E_R$ as the approximate ground state and energy of $H$.

The input state~$\ket{\Psi}$ and the size of $R$ highly affect the output quality of the QSCI method.
Since we select the $R$ most frequent bitstrings to create $\mathcal{S}_R$, the important bitstring to describe the exact ground state~$\ket{\psi_G}$ must appear with high probability by sampling of~$\ket{\Psi}$.
Therefore, the input state~$\ket{\Psi}$ should approximate the true ground state~$\ket{\psi_G}$ as closely as possible.
One of the reasonable choices for the input state is the state which has low energy expectation value~$\bra{\Psi}H\ket{\Psi}$ generated by VQE~\cite{kanno2023qsci}.
However, generating high-quality input states itself requires high computational cost.
Thus, various methods to generate input state have been proposed~\cite{nakagawa2024adapt-qsci}.
We have developed another approach for SQOA-QR and discuss it in Sec.~\ref{sec:linxfer}.

We diagonalize $H_R$ on the classical device, and this computational cost depends on $R$.
Thus, it is necessary to choose an $R$ that is small enough to be handled by a classical device.
However, since $R$ determines the size of $\mathcal{S}_R$, it requires appropriately large to capture the important bitstrings to describe the exact ground state~$\ket{\psi_G}$.
When $R=2^N$, the diagonalization of $H_R$ becomes equivalent to the diagonalization of $H$.
We assume $R_1 < R_2$, and write the corresponding energies as $E_{R_1}, E_{R_2}$, and the ground state energy of $H$ as $E_G$.
Then $E_{R_1}  \ge E_{R_2} \ge E_G$ holds.
This relationship indicates that the energy monotonically decreases as $R$ increases.

\section{LINXFER for QAOA-for-QRAO \label{sec:linxfer}}
We need to prepare the input state near the ground state of the quantum relaxed Hamiltonian to efficiently perform SQOA.
For this purpose, we demonstrate that LINXFER~\cite{Sakai2024-li}, a parameter setting heuristic for QAOA, also works effectively for QAOA-for-QRAO~\cite{He2025-br}.
In this section, we first explain the details of QAOA-for-QRAO and LINXFER and then present the experimental results.

\subsection{Details of LINXFER for QAOA-for-QRAO}
QAOA-for-QRAO is a method that applies the QAOAnsatz to a quantum relaxed Hamiltonian~\cite{He2025-br}.
The variational circuit of QAOA-for-QRAO is defined as
\begin{equation}
\ket{\psi(\bm{\beta}, \bm{\gamma})} = e^{-i\beta_p H_M}e^{-i\gamma_p \tilde{H}_C}\cdots e^{-i\beta_1 H_M}e^{-i\gamma_1 \tilde{H}_C}  \ket{\psi_0}
\label{eq:QAOA-for-QRAO}
\end{equation}
where $H_M$ is the mixer Hamiltonian and $\tilde{H}_C$ is the quantum relaxed Hamiltonian as defined in Eq.~\eqref{eq:relax_hamiltonian}.
This QAOAnsatz alternately applies the cost unitary $U_C(\gamma) = e^{-i\gamma \tilde{H}_C}$ and mixer unitary~$U_M(\beta) = e^{-i\beta H_M}$.
The design of the mixer Hamiltonian for the quantum relaxed Hamiltonian is not trivial.
He \textit{et al.}~\cite{He2025-br} considered simple single Pauli mixer Hamiltonians $H_M$, which apply the same Pauli operator to all qubits.
Specifically, they used one of three forms: $H_M^X = \sum_i X_i, H_M^Y = \sum_i Y_i$ or $H_M^Z = \sum_iZ_i$.
They investigated a parameter setting method leveraging parameter concentration, as initially considered for QAOA, and demonstrated that the method also worked effectively for QAOA-for-QRAO.
While more complex mixer Hamiltonians consisting of combinations of multiple Pauli operators are possible, we utilized these single Pauli mixer operators since our primary goal is to verify LINXFER's applicability to QAOA-for-QRAO.

Next, we explain LINXFER for QAOA-for-QRAO.
LINXFER was originally proposed as a parameter setting heuristic for QAOA that reduces the number of parameters to be optimized by restricting the parameters to linear form~\cite{Sakai2024-li}.
Despite this restriction, LINXFER shows parameter transferability.
This transferability allows parameters optimized on a smaller instance to be directly used for larger instances without re-optimization.
In LINXFER, we restrict the form of $\bm{\gamma}$ and $\bm{\beta}$ as
\begin{equation}
\begin{split}
    &\gamma_l = \gamma_\mathrm{slope}\frac{l}{p} + \gamma_\mathrm{int}\\
    &\beta_l = \beta_\mathrm{slope}\frac{l}{p} + \beta_\mathrm{int}.\\
\end{split}
\end{equation}
Here, $\tilde{\bm{\theta}} = (\gamma_\mathrm{slope},\gamma_\mathrm{int},\beta_\mathrm{slope}, \beta_\mathrm{int})$ represents the hyperparameters in LINXFER.
This restriction reduces the parameters in QAOA-for-QRAO from $2p$ to $4$.
For brevity, we denote parameterized states with parameters~$\tilde{\bm{\theta}}$ as~$\ket{\psi(\tilde{\bm{\theta}})}$.
First, we optimize the parameters~$\tilde{\bm{\theta}}$ to minimize the energy expectation value for a small problem. 
Then, we transfer the best parameters~$\tilde{\bm{\theta}}^*$ from this small problem to a larger problem.
In previous research, the parameters~$\tilde{\bm{\theta}}$ were optimized using Bayesian optimization~\cite{Sakai2024-li}.
For larger problems, we prepare a quantum state using these transferred parameters~$\tilde{\bm{\theta}}^*$, then apply a Rounding method to obtain a classical solution.
Following this procedure, we need to optimize only four parameters for a small problem.
This procedure can significantly reduce the computational cost compared to directly optimizing large problems.
We can also consider a parameter setting heuristic that utilizes these transferred parameters~$\tilde{\bm{\theta}}^*$ as the warm start initial value for variational parameter optimization.
We present the results of this method in the Appendix.

\subsection{Experimental Result}
\begin{figure}[t]
    \centering
    \includegraphics[width=\linewidth]{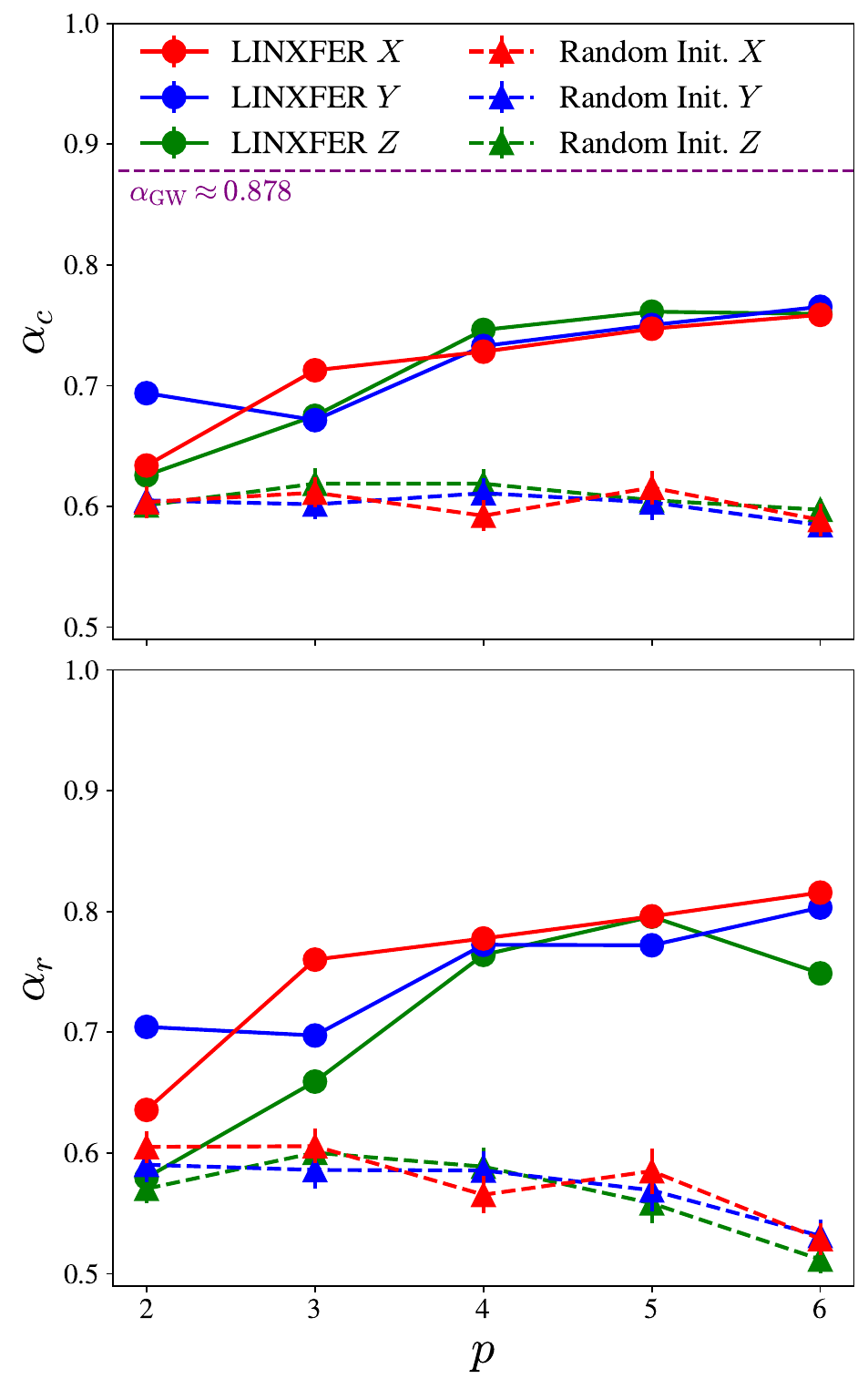}
    \caption{$p$-dependence of $\alpha_c$ (upper) and $\alpha_r$ (bottom) for $n=40$ using LINXFER (solid line) and Random Initialization (dotted line).
    Here, we varied $p$ from $2$ to $6$.
    Each color represents different mixers. }
    \label{fig:p-dependence}
\end{figure}

\begin{figure}[t]
    \centering
    \includegraphics[width=\linewidth]{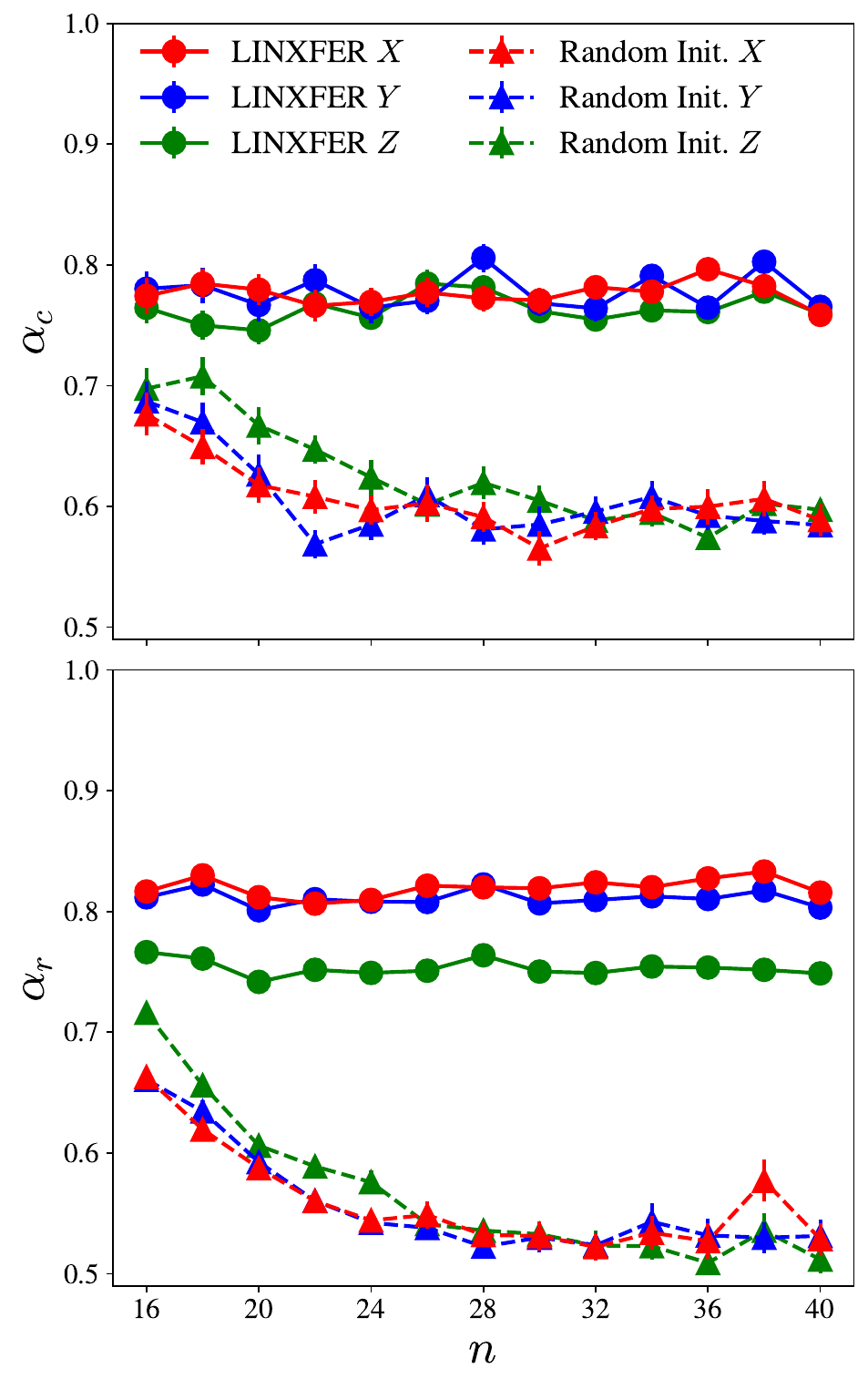}
    \caption{$n$-dependence of $\alpha_c$ (upper) and $\alpha_r$ (bottom) for $p=6$ using LINXFER (solid line) and Random Initialization (dotted line).
    Here, we varied $n$ from $16$ to $40$.
    Each color represents different mixers. }
    \label{fig:n-dependence}
\end{figure}
In this study, we evaluated the performance of LINXFER for $H_M^X, H_M^Y$ and $H_M^Z$.
For each of the mixer operators~$U_M(\beta) = e^{-i\beta H_M}$, we used the initial states of $\ket{+}^{\otimes N}, \ket{i}^{\otimes N}$, and $\ket{0}^{\otimes N}$, respectively.
In this study, we dealt with the unweighted MaxCut problem on 3-regular graphs.
We searched the best parameter, $\tilde{\bm{\theta}}^*$ using a 20-node problem and evaluated the parameter transferability of LINXFER using up to $16$ to $40$ nodes problem.
This parameter~$\tilde{\bm{\theta}}^*$ is calculated by Bayesian optimization using Optuna~\cite{optuna2019} independently for each mixer and $p$.
We define two approximation ratios as performance evaluation metrics:
\begin{equation}
    \alpha_r = \frac{\bra{\psi}\tilde{H}_C\ket{\psi}}{E_\mathrm{min}},\quad \alpha_c = \frac{C(\bm{s})}{C_\mathrm{min}}
\end{equation}
where $E_\mathrm{min}$ is the exact ground state energy of the quantum relaxed Hamiltonian $\tilde{H}_C$ calculated by direct diagonalization, and $\bra{\psi}\tilde{H}_C\ket{\psi}$ is the expectation value of the energy.
Similarly, $C_\mathrm{min}$ is the optimal value for the MaxCut problem, and $C(\bm{s})$ is the objective value obtained using Pauli Rounding.
We calculated $C_\mathrm{min}$ using Glover-Woolsey method~\cite{Glover1973further, Punnen2022exactalg}.
In order to investigate the performance of the simple QAOA-for-QRAO itself, we also experimented with QAOA-for-QRAO using random parameter initialization.
In this setting, we optimized the variational parameters from random initial values using COBYLA~\cite{Powell1994cobyla}.
In this study, we call this method Random Initialization~(Random Init.).
We took $50$ samples for each parameter setting.
The error bars in all plots represent standard errors.
We used Qamomile\footnote{https://github.com/Jij-Inc/Qamomile} and Quri-Parts\footnote{https://github.com/QunaSys/quri-parts} for quantum algorithm implementation.

\begin{algorithm}[t]
\caption{SQOA-QR : Sampling-based Quantum Optimization Algorithm with Quantum Relaxation}
\begin{algorithmic}[1]
\Procedure{SQOA-QR}{$Q$ : QUBO Problem, $R$ : Subspace size}
    
    \State $\tilde{H}_C \leftarrow$ \textsc{Quantum Relaxation}($Q$) 
    \Comment{Transform QUBO to quantum relaxed Hamiltonian}
    
    \State $U(\bm{\beta},\bm{\gamma}) \leftarrow$ \textsc{GenerateQAOAnsatz}($\tilde{H}_C$) 
    \Comment{Generate QAOAnsatz for Quantum relaxed Hamiltonian}
    
    \State $\bm{\beta}^*,\bm{\gamma}^* \leftarrow$ \textsc{SetLINXFERParameters}() 
    \Comment{Set precomputed LINXFER parameters}

    \State $\ket{\psi_R}, E_R \leftarrow$ \textsc{QSCI}($U(\bm{\beta}^*,\bm{\gamma}^*),\tilde{H}_C, R$) 
    \Comment{Estimate ground state and its Energy of $\tilde{H}_C$ using QSCI}
    
    \State $\bm{x} \leftarrow$ \textsc{PauliRounding}($\ket{\psi_R}$) 
    \Comment{Apply Pauli Rounding to $\ket{\psi_R}$ to obtain the classical solution}
    
    \State \Return $\bm{x}$
\EndProcedure
\end{algorithmic}
\label{alg:sqoa}
\end{algorithm}

First, we investigated whether LINXFER could work on QAOA-for-QRAO.
Figure~\ref{fig:p-dependence} shows the $p$-dependence of $\alpha_c$ and $\alpha_r$ for $n=40$.
Here, the color of each line indicates the difference in the mixer operators.
In the case of Random Initialization (dotted lines), the behavior of both $\alpha_c$ and $\alpha_r$ do not depend on the type of the mixer operator.
$\alpha_c$ is almost $0.6$ regardless of $p$, however $\alpha_r$ decreases slightly as $p$ increases.
Since Pauli Rounding is a heuristic rounding method, this result indicates that the slight decrease in the quality of the obtained quantum state does not have a significant impact.
This result suggests that in the case of Random Initialization, obtaining low energy quantum states becomes more difficult with increasing $p$.
In contrast, in the case of LINXFER (solid line), $\alpha_r$ and $\alpha_c$ increase with increasing $p$.
$\alpha_r$ with Pauli $X$ mixer achieve $\alpha_r = 0.8$ and $\alpha_c$ achieve more than $0.7$ at $p=6$.
In the case $\alpha_r$, $X$-mixer provides a better result than the other mixers.
However, for $\alpha_c$, there was no significant difference in the $p$-dependence of the mixers.
These results suggest that LINXFER for QAOA-for-QRAO works well on the $3$-regular MaxCut problem.
The purple dotted line is the guaranteed approximation ratio of the Goemans-Williamson algorithm, $\alpha_\mathrm{GW}\approx 0.878$~\cite{goemans1995improved, Williamson2011text}.
The performance of LINXFER does not achieve $\alpha_\mathrm{GW}$ in all $p$.

In Fig.~\ref{fig:n-dependence}, we plot $n$-dependence of LINXFER and Random Initialization for $p=6$.
Here, we transfer $\tilde{\bm{\theta}}^*$, which obtained a $20$-node problem to the problem up to $16$ to $40$ nodes.
In the case of Random Initialization, both $\alpha_c$ and $\alpha_r$ decrease as $n$ increases.
When $n$ is small, the $Z$-mixer provides better results on both $\alpha_c$ and $\alpha_r$.
This result is consistent with the previous study that does not assume a linear form~\cite{He2025-br}.
On the other hand, in the case of LINXFER, the $n$-dependence of both $\alpha_c$ and $\alpha_r$ are small.
In contrast to Random Initialization, $\alpha_r$ for $Z$-mixer is slightly lower than the other mixers.
Since the initial state for the $Z$-mixer is a classical state $\ket{0}$, there is a possibility that the Hilbert space is not fully explored due to the parameter restrictions imposed by LINXFER.
However, this difference in $\alpha_r$ does not affect the quality of the solution after Pauli Rounding.

We note that the quality of the solution obtained by LINXFER is not better than the results obtained using hardware-efficient ansatz in previous studies~\cite{Fuller2021-su, Teramoto2023-jd}.
However, the advantage of LINXFER for QAOA-for-QRAO is that it greatly reduces the computational cost for variational parameter optimization compared to the hardware-efficient approach.
We only need to optimize a much smaller number of parameters on a small instance, eliminating the need for parameter optimization on larger instances.
In the next section, we will discuss the quality of the solution when we use the state prepared by LINXFER as the input state of QSCI.

\section{Sampling-based Quantum Optimization Algorithm with Quantum Relaxation \label{sec:qsci}}

\begin{figure}[t]
    \centering
    \includegraphics[width=\linewidth]{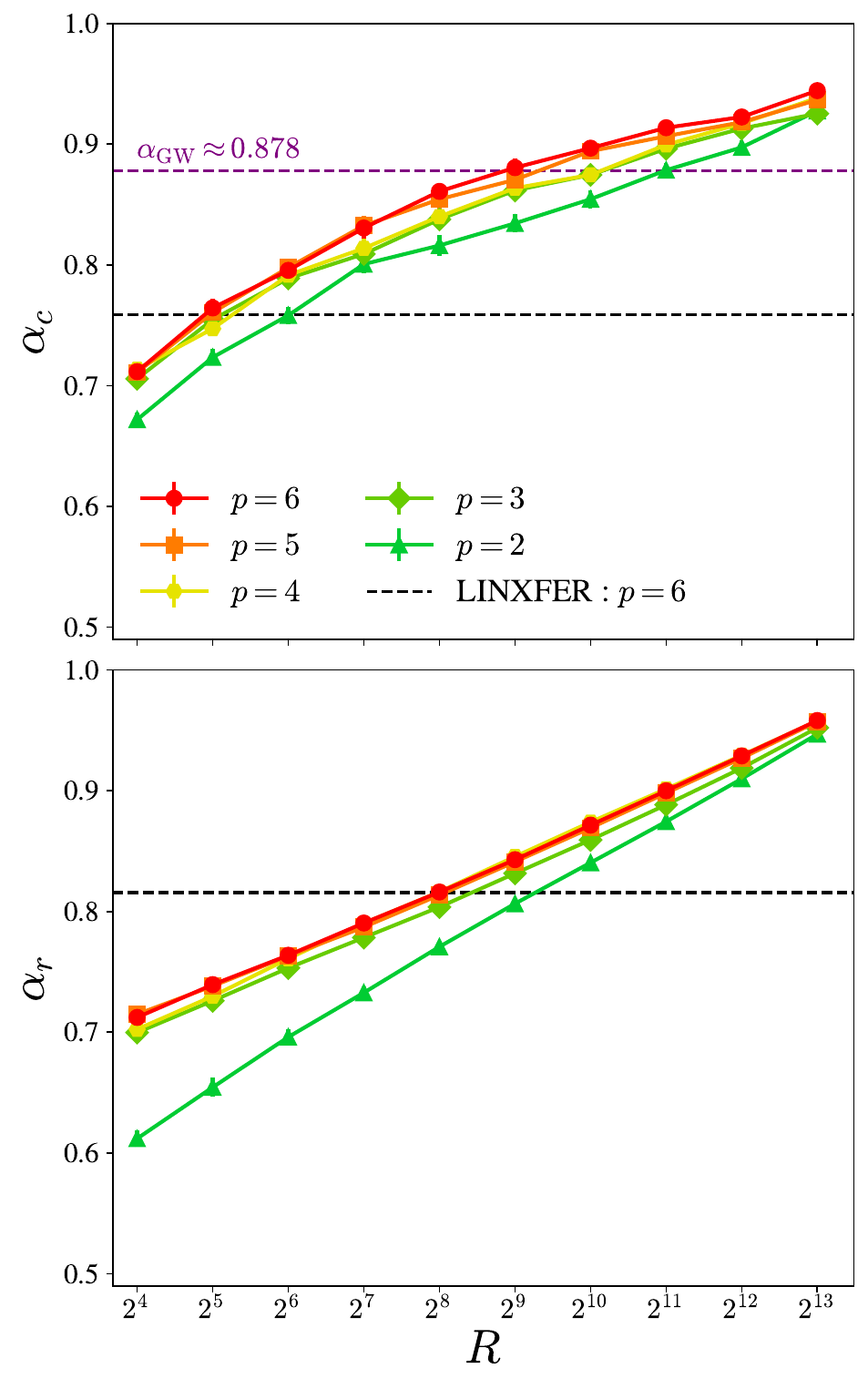}
    \caption{The $R$-dependence of $\alpha_c$ (upper) and $\alpha_r$ (bottom) for $n=40$  using SQOA-QR with $X$-mixer.
    We plot $p$ from $2$ to $6$ and $R$ from $2^4$ to $2^{13}$.
    Each color represents a different value of $p$.
    We plot the LINXFER result with $p=6$ (dashed line) as a baseline.
    As $p$ and $R$ increase, $\alpha_c$ and $\alpha_r$ increase.
    }
    \label{fig:R-dependence}
\end{figure}

\begin{figure}[t]
    \centering
    \includegraphics[width=\linewidth]{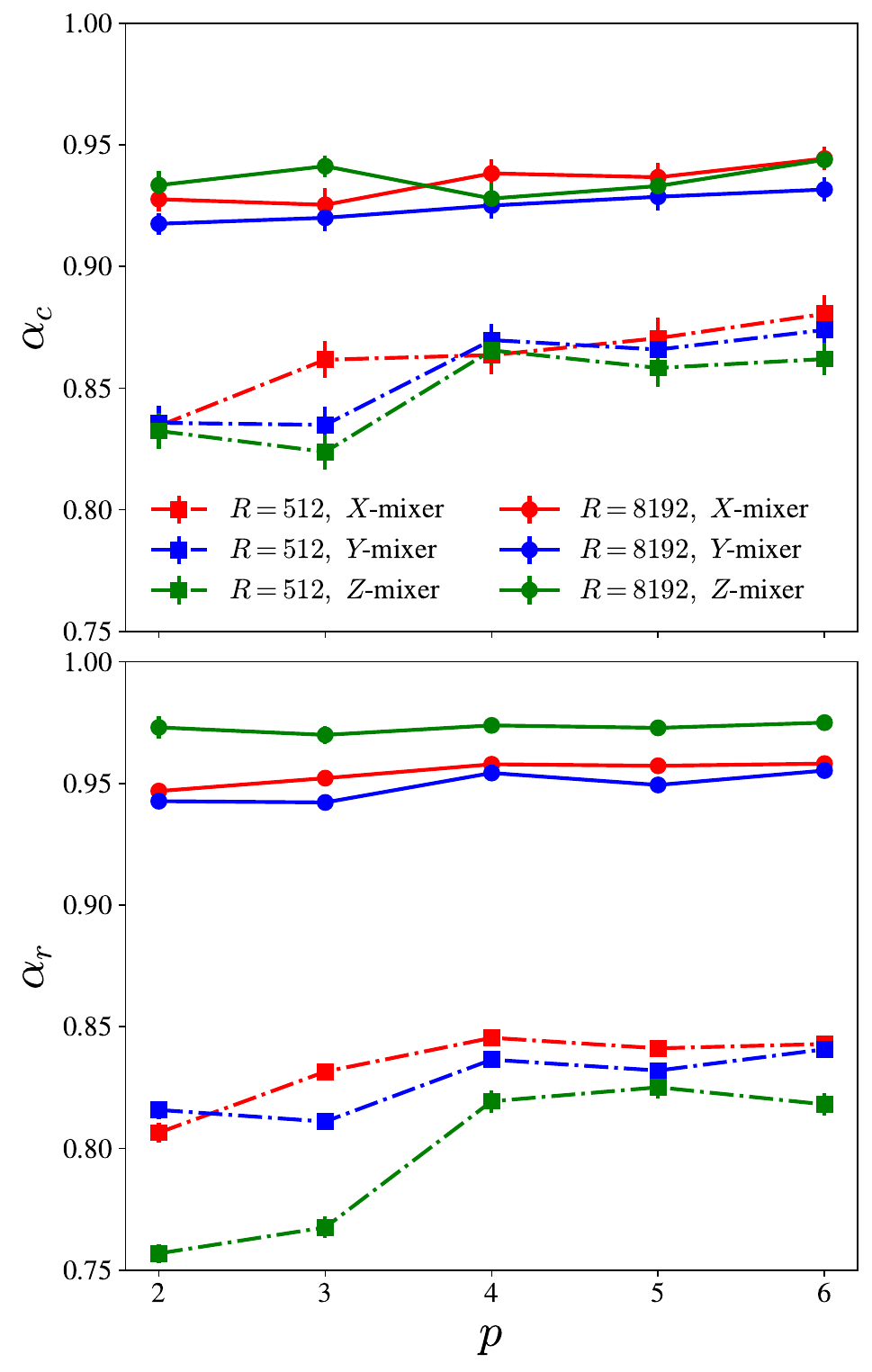}
    \caption{$p$-dependence of $\alpha_c$ (upper) and $\alpha_r$ (bottom) for $n=40$ using SQOA-QR.
    Here, we plotted $p$ from $2$ to $6$ and $R = 512, 8192$.
    Each color represents different mixers. 
    }
    \label{fig:p-dependence_qsci}
\end{figure}

\begin{figure}[t]
    \centering
    \includegraphics[width=\linewidth]{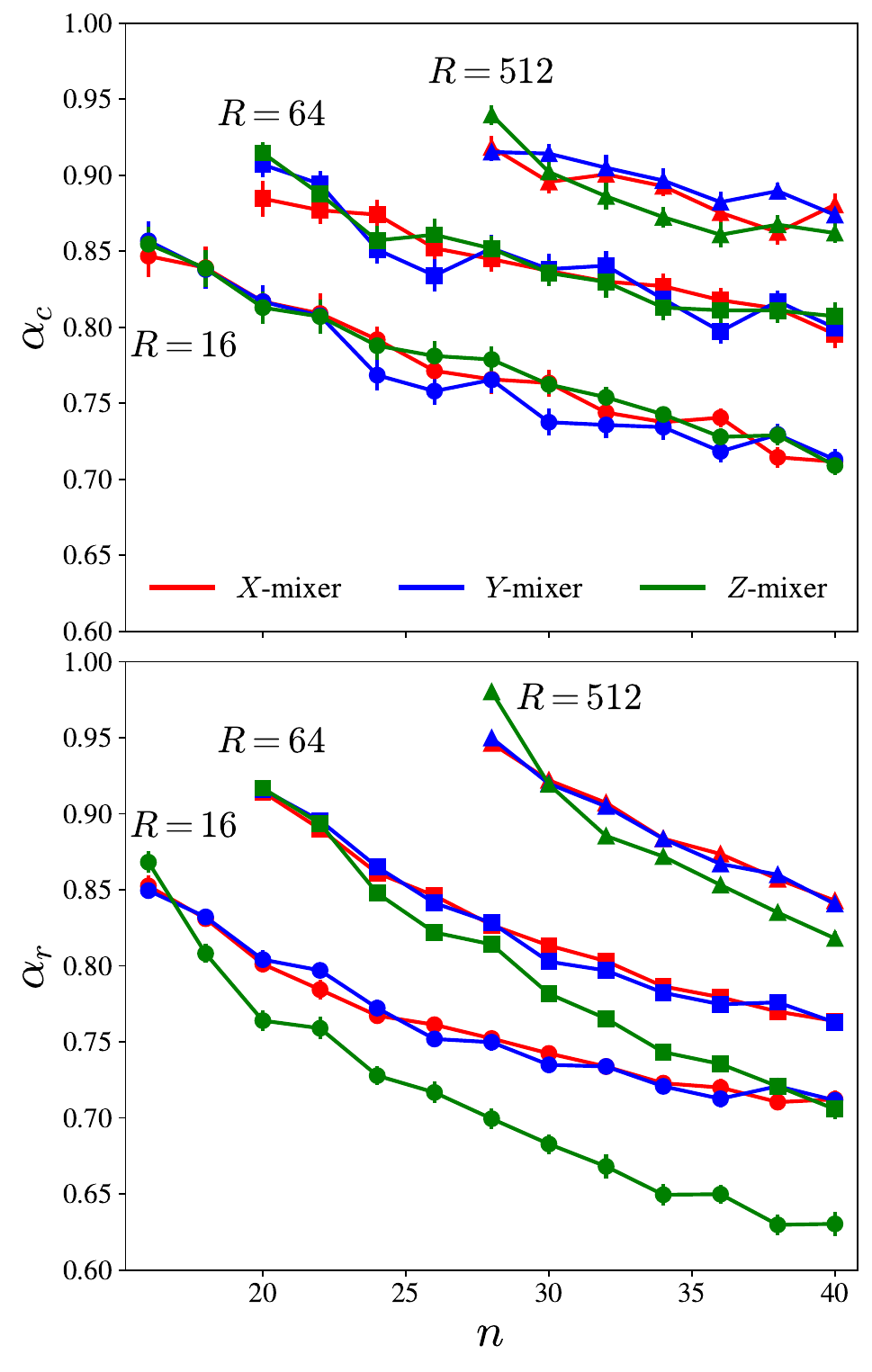}
    \caption{$n$-dependence of $\alpha_c$ (upper) and $\alpha_r$ (bottom) for $p=6$ using SQOA-QR.
    Here, we plotted $n$ from $16$ to $40$ and $R = 16, 64, 512$.
    Each color represents different mixers. 
    }
    \label{fig:n-dependence_qsci}
\end{figure}

This section explains the Sampling-based Quantum Optimization Algorithm with Quantum Relaxation (SQOA-QR) and describes the experimental results obtained by applying it to solve the 3-regular MaxCut problem.
Algorithm~\ref{alg:sqoa} shows the algorithm flow of SQOA-QR.
The inputs are the objective function of the QUBO problem to be solved and the parameter $R$, which is the size of the effective Hamiltonian in Eq.~\eqref{eq:effective_hamiltonian} and affects the accuracy of QSCI.
Increasing $R$ improves accuracy but also increases the computational cost because $R$ determines the size of the matrix to be diagonalized by the classical device.
Therefore, we must set $R$ appropriately according to the problem and the desired solution accuracy.
We discuss the impact of changing $R$ on SQOA-QR later.
SQOA-QR utilizes LINXFER, which is explained in Sec.~\ref{sec:linxfer}, to prepare the input state of QSCI (lines 2 to 4).
By precomputing the parameters of LINXFER on a small instance, we can prepare a QAOAnsatz circuit $U(\bm{\beta}^*,\bm{\gamma}^*)$ without variational parameter optimization.
Here, $\bm{\beta}^*,\bm{\gamma}^*$ are calculated from the LINXFER parameters $\tilde{\bm{\theta}}^*$.
Then, we can execute QSCI with the input state generated by $U(\bm{\beta}^*,\bm{\gamma}^*)$ to evaluate the ground state~$\ket{\psi_R}$ and its energy $E_R$ of $\tilde{H}_C$ (line 5).
Finally, we apply Pauli Rounding to the obtained state~$\ket{\psi_R}$ to decode the classical solution $\bm{x} \in \{0,1\}^n$ (line 6).
SQOA-QR outputs this classical solution as the final result (line 7).

In our experiments, we evaluated the 3-regular MaxCut problems.
In QSCI, we used the same parameters~$\tilde{\bm{\theta}}^*$ as in Sec.~\ref{sec:linxfer} for state preparation and measured the input state $10^6$ times to obtain bitstrings.
When the problem is encoded in an $n_q$-qubit quantum relaxed Hamiltonian, we performed diagonalization of the effective Hamiltonian up to $R=2^{n_q - 1}$.
We collected 50 independent samples for each combination of mixer operators, $p$ and $R$.

First, we investigated the $R$ dependence of the approximation ratios $\alpha_c$ and $\alpha_r$.
Figure~\ref{fig:R-dependence} demonstrates the results for the 40-node instances with the $X$-mixer.
Each solid line represents the QSCI results for different $p$, and the black and purple dotted lines represent the LINXFER results for $p=6$ and the approximation ratio of the Goemans-Williamson algorithm, respectively, as a baseline.
Since the ground state of $\tilde{H}_C$ is evaluated by QSCI, it is clear that $\alpha_r$ monotonically increases with $R$.
On the other hand, although Pauli Rounding is a heuristic rounding algorithm, $\alpha_c$ also monotonically increases with $R$.
This suggests that obtaining a quantum state that is closer to the ground state provides a better classical solution.
Moreover, we achieved $\alpha_c \geq \alpha_\mathrm{GW}$ when $p = 6$ and $R = 2^{9}$.
This result demonstrates that we can obtain high-quality classical solutions even when we select a relatively small $R$ compared to the total number of computational basis states.
Since we can encode almost all 40-node problems using $14$ qubits, $\alpha_r$ achieves a high approximation ratio regardless of the choice of $p$ with $R=2^{13}$.

In the case of $p=2$, as we have seen in Sec.~\ref{sec:linxfer}, the quality of the input state is not high enough, so the approximation ratio of QSCI is also not high at smaller $R$.
In contrast, in the case of $p\geq 3$, the quality of the input state improves, so the required $R$ becomes smaller.
For $\alpha_r$, we found that when $p \geq 3$, only about half the $R$ value is required to exceed the approximation ratio of LINXFER with $p=6$ compared to when $p=2$.
Moreover, in the case of $\alpha_c$, the LINXFER accuracy is exceeded with smaller $R$ than that required for $\alpha_r$ across all $p$.
However, we observed a similar pattern for both approximation ratios: when $p\geq 3$, only about half of the $R$ value is required to exceed the approximation ratio of LINXFER compared to when $p=2$.
Therefore, it is possible to reduce the computational cost of QSCI by using a better input state.
In other words, at fixed $R$, solution quality can be improved by using a better input state. 

Next, we investigated the $p$-dependence of SQOA-QR.
Figure~\ref{fig:p-dependence_qsci} demonstrates the $p$-dependence of SQOA-QR for $R=512$ and $R=8192$.
Each color represents a different type of mixer operator.
In the case of $R=512$, we can see that $\alpha_r$ and $\alpha_c$ increases as $p$ increases.
These behaviors resemble what we observed in LINXFER (Fig.~\ref{fig:p-dependence}), although the approximation ratios increase more gradually than in the LINXFER case.
On the other hand, in the case of $R=8192$, there is almost no dependence on $p$ in both $\alpha_r$ and $\alpha_c$.
As discussed in Fig.~\ref{fig:R-dependence}, in the case of $R=8192$, we have already selected a sufficiently large $R$, so the effect of the quality of the input state is almost negligible.
Interestingly, in the case of $R=8192$, the $Z$-mixer provides the highest quality solution for $\alpha_r$.
This result contrasts with the case of $R=512$, where the $Z$-mixer performs worse than all other mixer operators.

Finally, we investigated the problem size dependence of SQOA-QR as shown in Fig.~\ref{fig:n-dependence_qsci}.
Here, the maximum value of $R$ depends on the number of required qubits $n_q$ to encode the problem, and $R$ is limited to $2^{n_q}$.
Therefore, we selected $R=16, 64, 512$ and plotted the results for $p=6$ in Fig.~\ref{fig:n-dependence_qsci}.
Here, both $\alpha_c$ and $\alpha_r$ decrease as problem size $n$ increases.
This behavior occurs because as $n$ increases, the maximum possible value of $R$ becomes larger, making our fixed plotted $R$ relatively smaller.
Interestingly, the $Z$-mixer decreases much faster in $\alpha_r$ as $n$ increases than the $X, Y$-mixers.
This result reflects that, as seen in Fig.~\ref{fig:n-dependence}, the $Z$-mixer has a lower $\alpha_r$ than the $X, Y$-mixers in LINXFER.
On the other hand, such mixer-dependent behavior is hardly seen in the case of $\alpha_c$.
This behavior is also consistent with the results shown in Fig.~\ref{fig:n-dependence}.

\section{Summary \label{sec:summary}}
This paper introduces the Sampling-based Quantum Optimization Algorithm (SQOA) as a novel approach to combinatorial optimization.
SQOA operates through a three-stage process: first, the target problem is encoded into a non-diagonal quantum Hamiltonian. Subsequently, Quantum-Selected Configuration Interaction (QSCI) is employed to estimate the ground state and its corresponding energy. Finally, classical solutions are extracted from the obtained quantum state using a rounding algorithm. The efficacy of SQOA hinges on two crucial elements: the method of problem encoding and the preparation of the input state for QSCI.

For our encoding approach, we utilized the quantum relaxation method to encode the optimization problem into a Hamiltonian~\cite{Fuller2021-su}.
This method achieves a compression ratio three times higher than traditional Ising Hamiltonian approaches.
For state preparation, we demonstrated that LINXFER~\cite{Sakai2024-li}, which is proposed for QAOA, also works for QAOA-for-QRAO when applied to the 3-regular MaxCut problem.
LINXFER simplifies the parameter landscape by constraining QAOA parameters to a linear form, thereby reducing the number of parameters from $2p$ to a mere four.
In Sec.~\ref{sec:linxfer}, we demonstrate that even with these restricted parameters, LINXFER for QAOA-for-QRAO exhibits transferability across various sizes of 3-regular MaxCut instances. Our results show that the best variational parameters obtained using a single 20-node instance were successfully transferred to 40-node instances, providing better approximation ratios than Random Initialization.

We implemented SQOA-QR, which uses LINXFER for QAOA-for-QRAO to prepare the input state for QSCI.
In Sec.~\ref{sec:qsci}, we demonstrated the experimental results of SQOA-QR on the 3-regular MaxCut problem.
The size of subspace $R$ and the number of repetition layers $p$ in QAOA-for-QRAO are important parameters in SQOA-QR.
We show in Fig.~\ref{fig:R-dependence} that as $R$ and $p$ increase, both approximation ratios $\alpha_r$ and $\alpha_c$ increase.
Even with the moderate size of $R$, we obtain high-quality classical solutions with $\alpha_c \geq \alpha_\mathrm{GW} \approx 0.878$.
At smaller $R$ values, the behavior of approximation ratios is similar to LINXFER, so as $p$ increases, approximation ratios increase.
However, if we set larger $R$ values, these approximation ratios show almost no dependency on $p$.
Finally, we examined the $n$-dependence of $\alpha_r$ and $\alpha_c$ using SQOA-QR.
We found that $\alpha_r$ with the $Z$-mixer decreases more rapidly than the other mixers as $n$ increases.
However, we could not find this behavior for $\alpha_c$.
Therefore, the choice of the mixer Hamiltonian has almost no effect on the quality of the classical solution.

In our experiments, we only dealt with the 3-regular MaxCut problem.
However, both SQOA-QR and  LINXFER for QAOA-for-QRAO are not limited to this problem.
Evaluating the performance of both methods on different optimization problems remains important future work.
Also, since SQOA is a sampling-based algorithm, it is expected to be robust against noise.
However, the cost operator in QAOA-for-QRAO is not hardware-efficient.
Therefore, the effectiveness of SQOA-QR on real quantum devices requires further investigation.

SQOA-QR provided high-quality solutions without optimizing variational parameters for large instances by obtaining optimal parameters for a small instance.
Recently, several algorithms such as Recursive QRAO~\cite{Kondo2025-dd} and QR-BnB~\cite{matsuyama2024qr-bnb} have been proposed that repeatedly perform QRAO to obtain high-quality solutions for large instances.
Applying SQOA-QR as a subroutine for these more complex algorithms is an interesting future direction.
Since it is possible to eliminate the optimization loop for the variational parameters, there is potential for significantly speeding up these algorithms.

Finally, although this study employed Quantum Relaxation for constructing the problem Hamiltonian, the SQOA framework is inherently flexible and not restricted to this specific method. Exploring alternative Hamiltonian constructions, such as the quantum Hamiltonian version of Pauli-correlation encoding~\cite{sciorilli2024largescale}, as candidates for SQOA represents another compelling direction for future research. Ultimately, the continued development of suitable cost Hamiltonians and their integration within the SQOA framework holds the key to unlocking new frontiers in this field.

\section*{Acknowledgment}
H.M. thanks for the fruitful discussion with Keita Kanno and Toru Shibamiya.

\appendix[Parameter Setting Heuristic using fine-tuning]
\begin{figure}[tb]
    \centering
    \includegraphics[width=\linewidth]{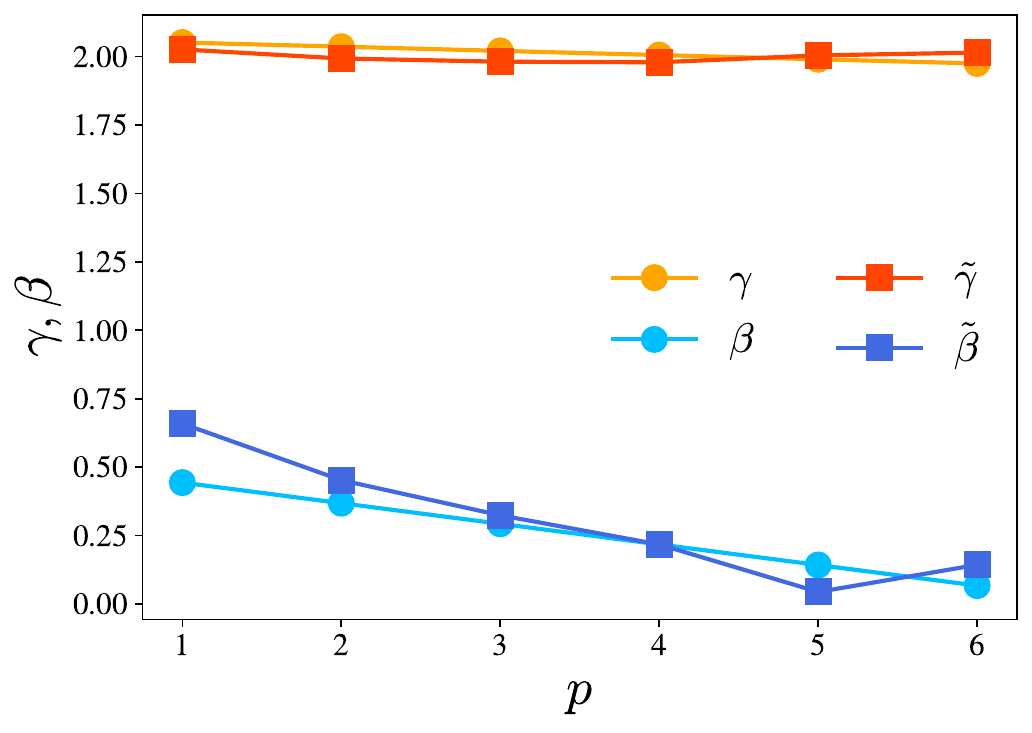}
    \caption{Comparison of initial variational parameters $\bm{\gamma}, \bm{\beta}$ obtained by LINXFER and $\tilde{\bm{\gamma}}, \tilde{\bm{\beta}}$ after fine-tuning}
    \label{fig:reopt_parameter}
\end{figure}
\begin{figure}[t]
    \centering
    \includegraphics[width=\linewidth]{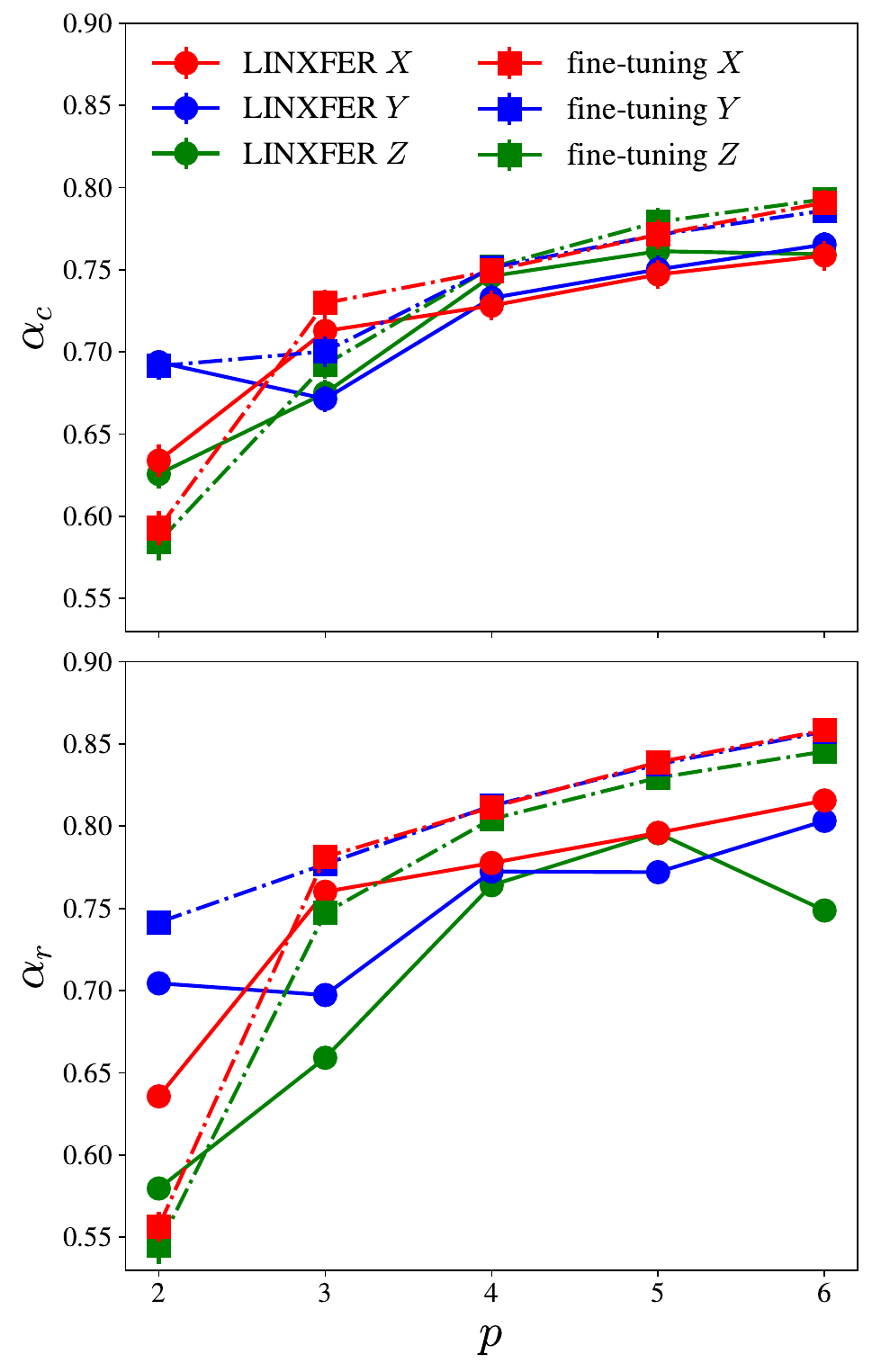}
    \caption{The $p$-dependence of $\alpha_c$ (upper) and $\alpha_r$ (bottom) for LINXFER (solid line) and fine-tuning (dash-dot line).}
    \label{fig:reopt_p}
\end{figure}

In the main text, we compared the performance of Random Initialization and LINXFER for QAOA-for-QRAO.
Another parameter setting heuristic worth exploring is using the best LINXFER parameters, $\tilde{\bm{\theta}}^*$, as warm start initial values at each instance. 
We refer to this parameter setting heuristic as fine-tuning.
This appendix investigates whether fine-tuning improves performance compared to directly using LINXFER parameters.
For the fine-tuning process, we prepared $\bm{\gamma}$ and $\bm{\beta}$ from $\tilde{\bm{\theta}}^*$ as the initial value of the optimization for the variational parameters.
Then, we optimized these parameters using COBYLA.
Unlike in LINXFER, we did not assume the linearity of the parameters during this optimization. 
We present the results for 40-node instances.

Figure~\ref{fig:reopt_parameter} shows both initial values of parameters~$\bm{\gamma}, \bm{\beta}$ used in the fine-tuning, and the resulting parameters~$\tilde{\bm{\gamma}}, \tilde{\bm{\beta}}$ after fine-tuning.
The shapes of $\bm{\gamma}$ and $\tilde{\bm{\gamma}}$ remain almost unchanged before and after fine-tuning.
In addition, although there is some deviation from linearity in the behavior of $\tilde{\bm{\beta}}$, there is no significant change in the overall trend.
This result suggests that the LINXFER parameters found in a $20$-node problem are relatively stable even for 40-node problems.

Next, Figure~\ref{fig:reopt_p} shows the results of investigating the change in the approximation ratios due to fine-tuning.
Here, the solid line represents results using LINXFER parameters, and the dash-dot line represents results after fine-tuning.
In the case of $p=2$, there are some cases where fine-tuning worsens both $\alpha_r$ and $\alpha_c$.
However, in the case of $p\geq3$, fine-tuning improves both $\alpha_r$ and $\alpha_c$.
The degree of improvement for $\alpha_r$ is slightly larger than for $\alpha_c$.
The overall $p$-dependence does not change significantly before and after fine-tuning.
However, for $p=6$, the values of both $\alpha_r$ and $\alpha_c$ obtained by fine-tuning heuristic converge to almost the same accuracy regardless of which mixer is selected.
Thus, the parameter fine-tuning heuristic using the LINXFER parameters works well.

\printbibliography[title=References]

\end{document}